\documentclass[aps,prb,twocolumn]{revtex4-2}
\usepackage[utf8]{inputenc}
\usepackage{fontenc}
\usepackage{amsmath,amssymb}
\usepackage{graphicx}
\usepackage{textcomp}
\usepackage{mathtools}
\usepackage{bm}
\usepackage[section]{placeins}
\usepackage{placeins}

\begin{document}

  \begin{abstract}
  	Magnetic skyrmions with its topologically protected, nano-sized spin
  	textures have already earned immense fame as information carriers
  	due to their stability and low-current mobility. While ferromagnetic
  	skyrmions suffer from a transverse deflection ($i.e.,$ the skyrmion Hall effect), their antiferromagnetic counterparts promise straight-line motion and ultrafast dynamics. Here we present a numerical study of the dynamics of lattice-based antiferromagnetic skyrmions driven by spin-transfer torque for which the Landau-Lifshitz-Gilbert-Slonczewski (LLGS) equation is solved using a fourth-order Range-Kutta integration.
  	Then we conduct a detailed Monte Carlo study of the two-dimensional classical XY model to quantify how spatial anisotropy and Dzyaloshinskii–Moriya (DM) coupling reshape its thermal response across multiple lattice sizes.  By tuning the ratio $J_y/J_x$ from 0.2 to 6.0, we document a systematic evolution of the specific-heat anomaly. For example, in the quasi‑one‑dimensional limit, $C_V$ exhibits a broad, low-temperature hump, whereas stronger anisotropy yields sharper peaks that migrate to higher inverse temperatures.  Finite-size scaling of peak heights versus $[\ln~L]$ and pseudo-critical couplings against $1/[\ln~L]^2$ confirms the crossover from quasi‑1D fluctuations to a two-dimensional Kosterlitz–Thouless transition.  Incorporating a DM interaction further enriches this landscape. At $D/J=0.1$, the main peak shifts modestly upward and is slightly suppressed; raising $D/J$ elevates the peak magnitudes and creates a pronounced low-temperature plateau.  This residual $C_V$ signals enduring chiral excitations and complex spin-twist textures beyond simple vortex unbinding.  Our findings chart how directional and chiral couplings can be harnessed to tune pseudo-critical temperatures and thermodynamic signatures in two-dimensional magnets, providing a practical blueprint for engineering topological spin systems.
  	
  \end{abstract}

\title{Reviewing Current‐Driven Dynamics and Monte-Carlo based Analysis of Thermodynamic Properties of a Magnetic Skyrmion Crystal}
\author{Rajdip Banerjee}\email{rajdip.banerjee9696@gmail.com}
\affiliation{Department of Physics, Indian Institute of Engineering Science and Technology , Howrah, West Bengal-711103, India}
\author{Satyaki Kar}\email{satyaki.phys@gmail.com}
\affiliation{Department of Physics, A.K.P.C. Mahavidyalaya, Bengai, West Bengal-712611, India}
	  \maketitle
\date{\today}
	\section{Introduction}
	Topological solitons in field theory, first proposed by Skyrme, manifest in magnets as magnetic skyrmions: whirl‐like spin textures arise from the competition between symmetric Heisenberg exchange and antisymmetric Dzyaloshinskii–Moriya interactions (DMI) in noncentrosymmetric magnets.Originally predicted by Bogdanov and Yablonskii as thermodynamically stable vortices in chiral magnets \cite{1}, skyrmion crystals have attracted intense interest for their nanoscale size, robustness, and efficient current‐driven dynamics \cite{2}.However, most studies to date have focused on ferromagnetic hosts, where the net magnetization gives rise to stray fields\cite{3} and limits skyrmion velocity via the skyrmion Hall effect.Antiferromagnetic (AFM) skyrmions, in which two sublattice magnetizations are equal in magnitude but opposite in direction, offer key advantages: absence of net magnetization suppresses unwanted dipolar fields, and coupled dynamics eliminate the skyrmion Hall effect, promising faster, more controllable motion under spin‐torques \cite{4}.Pioneering theoretical work demonstrated that individual AFM skyrmions are stable against thermal fluctuations and can be nucleated and driven efficiently by spin currents \cite{5}.\\
	AFM Skyrmion formation is achieved by injecting a brief, perpendicularly oriented spin-polarized current into an antiferromagnetic nanodisk initially in its uniform ground state. During the pulse, spins within the targeted circular region are driven out of their equilibrium alignment. Upon cessation of the current, the disturbed spins relax into a closed, toroidal domain-wall configuration that simultaneously lowers the Dzyaloshinskii–Moriya and exchange energies, resulting in a well-defined skyrmion whose two magnetic sublattices carry opposite unit topological charges\cite{4}. Notably, the ultimate skyrmion diameter is governed solely by intrinsic material parameters—such as DMI strength and anisotropy—rendering it independent of both the injection-region size and pulse duration\cite{5}.Once created, antiferromagnetic skyrmions remain exceptionally robust at room temperature, owing to their zero net magnetization (which eliminates stray‐field–driven collapse) and the complete cancellation of the Magnus force—yielding perfectly straight motion under applied torques\cite{18}, free from the Hall‐angle deviations that bedevil ferromagnetic skyrmions\cite{4}\cite{5}.Magnetic skyrmions are characterized by an integer–valued topological charge (or winding number), which counts how many times the local magnetization field $\mathbf{m}(x,y)$ wraps the unit sphere:
	\begin{equation}
		Q \;=\;\frac{1}{4\pi}\int \! \mathrm{d}^2r\;\mathbf{m}\cdot\bigl(\partial_x\mathbf{m}\times\partial_y\mathbf{m}\bigr)\,,
	\end{equation}
	and which cannot change under any smooth deformation of the spin texture, providing a topology‐related energy barrier against collapse \cite{6}
	
	In systems with a homogeneous Dzyaloshinskii–Moriya interaction, isolated skyrmions almost always carry $|Q|=1$, corresponding to a single $2\pi$ twist of the spins from core to periphery.  Higher‐charge textures ($|Q|>1$) generally require additional interactions or magnetic frustration to stabilize \cite{2}.
	
	In antiferromagnetic materials, each of the two magnetic sublattices hosts a skyrmion with opposite unit charge, $Q_A=+1$ and $Q_B=-1$, so that the total charge vanishes.  This cancellation eliminates stray‐field interactions and the Magnus‐force–induced deflection under current drive, enabling perfectly straight motion and greatly reduced skyrmion–skyrmion repulsion \cite{7}.
	More recently, the concept has been extended to long range ordered skyrmion crystals (SkX) in AFM materials, opening new avenues for high‐density information storage and emergent spintronic functionalities.\cite{17}\\
	Building on our zero‐temperature analysis of skyrmion nucleation, we turn to classical Monte Carlo (MC) simulations to elucidate how thermal fluctuations modulate the formation, stability, and phase transitions\cite{12}\cite{13} of magnetic skyrmion crystals. In our study, we simulate a two‐dimensional xy model lattice (20×20 spins with periodic boundaries), running over 110000 Metropolis steps per spin at each temperature to guarantee thorough exploration of configuration space and capture rare,fluctuation‐driven events .From these equilibrated ensembles, we extract key thermodynamic observables—mean energy, specific heat (whose pronounced peaks pinpoint the Néel→SkX and SkX→spiral transitions), magnetic susceptibility etc.Unlike direct time‐integration techniques, Monte Carlo naturally incorporates entropic effects that are essential for stabilizing the skyrmion “A‐phase” and concentrates computational effort on the most relevant configurations through importance sampling . Its uncomplicated framework readily accommodates Heisenberg exchange, DMI, and uniaxial anisotropy, while providing both average values and higher‐order cumulants—tools that are indispensable for distinguishing first‐order from continuous transitions and for guiding the experimental realization of room‐temperature skyrmion materials.
	\section{AFM Skyrmion Spin Texture}
	In order to investigate the formation and stability of antiferromagnetic (AFM) skyrmions within a continuous‐field framework, we initialized a two‐dimensional Néel‐type skyrmion texture on a 60×60 grid and numerically integrated the coupled Landau–Lifshitz–Gilbert–Slonczewski equations using a classical fourth‐order Runge–Kutta scheme. The initial spin configuration is prescribed analytically to represent a Néel-type skyrmion, with the polar angle 
  $\theta(r)=\pi(1-r/R)$ for $r\leq R$,
	 ensuring a smooth 360° rotation of magnetization from the core to the boundary.Following the formalism established by Zhang et al.\cite{8}  , the total Hamiltonian is expressed as:
	 \begin{equation}
	 	H_{\text{AFM}}=\sum_{i,j}J\mathbf{m_i . m_j}+\sum_{i,j}\mathbf{D.(m_i \times m_j)}-\sum_{i}\mathbf{K}\mathbf{(m_i^z)}^2
   	\end{equation}
	Here, $\mathbf{m_i} \in \mathbb{R}^3$ is a unit vector representing the local magnetization direction at lattice site,i i.e $|\mathbf{m}_i| = 1$. Each term in this Hamiltonian corresponds to a distinct physical interaction. \\
	The first term inthe AFM hamiltonian($H_{\text{AFM}}$) is the antiferromagnetic exchange interaction.This term favours antiparallel alignment between neighboring spins, with 
	$J>0$ indicating an antiferromagnetic exchange constant. It minimizes energy when neighboring spins are oppositely oriented. This exchange interaction is the dominant term stabilizing the Néel ground state in collinear AFMs and contributes significantly to the stiffness of the spin texture.The second term in the hamiltonian is the interfacial Dzyaloshinskii–Moriya Interaction (DMI).The DMI arises due to broken inversion symmetry at the interface and strong spin–orbit coupling, leading to a preference for orthogonal (canted) spin configurations.For interfacial DMI, the DMI vector $\mathbf{D_{\text{ij}}}$	typically lies in the film plane and is perpendicular to the bond direction between sites $i$ and $j$. This term is crucial for stabilizing skyrmions: it penalizes collinear spin arrangements and energetically favors twisted configurations, such as the Néel-type skyrmion observed here. The third term in the hamiltonian, is perpendicular magnetic anisotropy (PMA).This term energetically favors alignment of the magnetization vector along the out-of-plane direction (±z axis). When 
	$K>0$, spins prefer to align either “up” or “down,”. PMA competes with the exchange and DMI terms, influencing the skyrmion radius. It is essential for maintaining a robust $\mathbf{m_z}$ profile across the skyrmion, preventing it from collapsing or diffusing laterally\cite{8}. 
	
	 \section{Effective Field and Spin Dynamics}The Landau-Lifshitz-Gilbert (LLG) and Landau-Lifshitz (LL) equations play an essential role for
	 explaining the dynamics of magnetization in solids\cite{9}.The dynamics of the AFM system are governed by the Landau–Lifshitz–Gilbert–
	 Slonczewski (LLGS) equation:\\
	 	\begin{equation}
	    	\mathbf{\frac{d\bm{m}_i}{dt}}
	 		\;=\;
	 		-\gamma\,\mathbf{\bm{m}_i \times \bm{H}_i^{\mathrm{eff}}}
	 		\;+\;
	 		\alpha\,\mathbf{\bm{m}_i \times \frac{d\bm{m}_i}{dt}}
	 		\;+\;
	 		\beta\,\mathbf{\bm{m}_i \times \bigl(\bm{m}_i \times \bm{p}\bigr)}
	 	\end{equation}
	 	where:
	 	\begin{itemize}
	 		\item $\gamma$ is the gyromagnetic ratio,
	 		\item $\alpha$ is the Gilbert damping constant,
	 		\item $\beta$ is the spin-transfer torque (STT) coefficient,
	 		\item $\bm{p}$ is the direction of spin polarization,
	 		\item $\mathbf{\bm{H}_i^{\mathrm{eff}} = -\partial H_{\mathrm{AFM}} / \partial \bm{m}_i}$ is the effective magnetic field derived from the Hamiltonian.
	 	\end{itemize}
	 	The Slonczewski like STT term governs how spin polarized currents couple to the spin system and is responsible for skyrmion nucleation and motion.current was injected perpendicular to the film plane (CPP configuration) with polarization direction $\bm{p} = -\hat{\bm{z}}$ to nucleate the skyrmion.
	 	
	 The dynamics are evolved using a fourth-order Runge–Kutta (RK4) integration of the Landau–Lifshitz–Gilbert (LLG) equation augmented by a Slonczewski-type spin-transfer torque (STT) term. After 5000 time steps, we analyze the stabilized spin configuration to extract key physical quantities—including the out-of-plane magnetization, in-plane texture, topological charge density,to characterize the resulting skyrmion and verify its topological and chiral properties\cite{16}.
	 
	 \begin{figure}[htbp]
	 	\centering
	 	\includegraphics[width=0.5\textwidth]{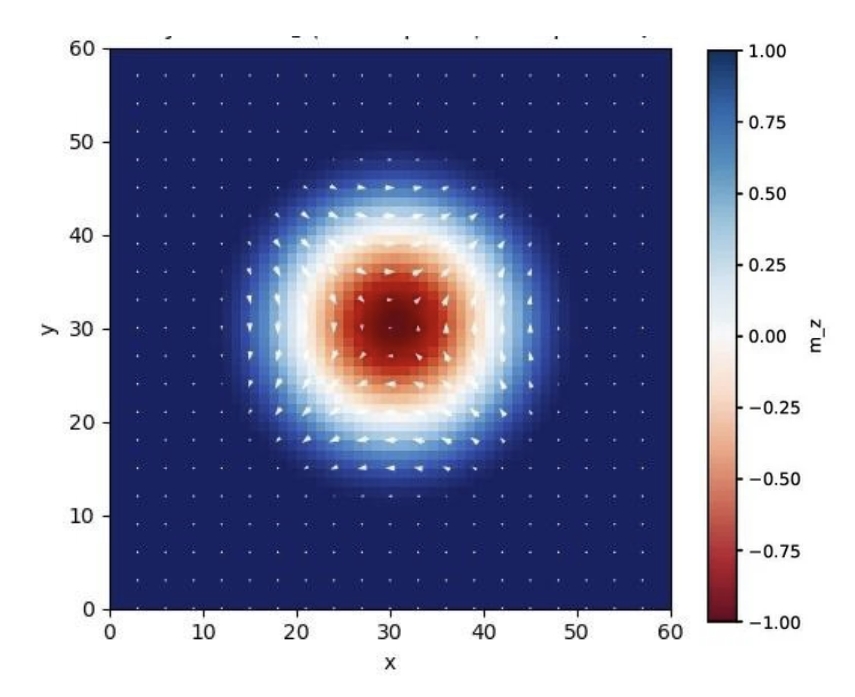} 
	 	\caption{Final antiferromagnetic skyrmion configuration on a $60 \times 60$ continuous-field grid. The background colormap represents the out-of-plane magnetization component $m_z$, with dark red ($\approx -1$) at the skyrmion core and dark blue ($\approx +1$) in the uniform surrounding region. Overlaid white arrows depict the in-plane spin vectors $(m_x, m_y)$, sampled on every third grid point for clarity. The radially symmetric pattern of these arrows confirms a Néel-type skyrmion, as expected for a texture stabilized by interfacial Dzyaloshinskii–Moriya interaction (DMI). Starting from an analytical ansatz $\theta(r) = \pi \left(1 - \tfrac{r}{R}\right)$, the system was evolved using fourth-order Runge–Kutta integration of the Landau–Lifshitz–Gilbert–Slonczewski (LLGS) equation for 5000 steps ($\Delta t = 0.55$), with parameters: exchange constant $J = 20 \times 10^{-21}\,\mathrm{J}$, DMI strength $D = 10 \times 10^{-21}\,\mathrm{J}$, uniaxial anisotropy $K = 2 \times 10^{-21}\,\mathrm{J}$, and damping $\alpha = 0.3$. The resulting structure exhibits a skyrmion with core polarity $m_z \approx -1$ and a surrounding domain-wall region where $m_z \approx 0$. (Colorbar range: $-1 \le m_z \le +1$.)}
	 	\label{fig:example}
	 \end{figure}	
	 
	  \begin{figure}[!ht]
	  	\centering
	  	\includegraphics[width=\linewidth]{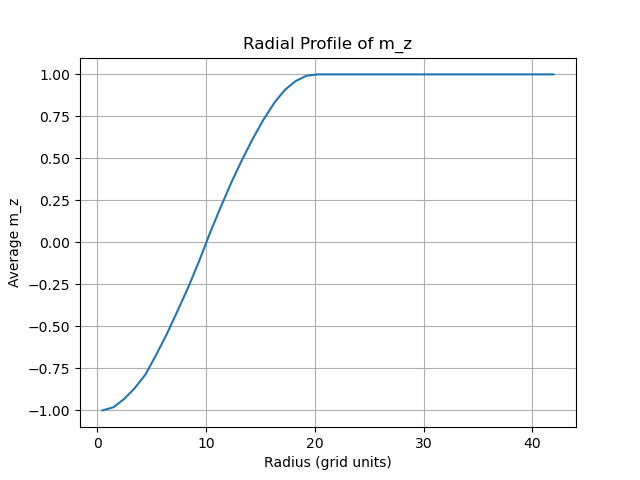}
	  	\caption{Radial profile of $m_z$ as a function of radial distance $r$.}
	  	\label{fig:radial}
	  \end{figure}
	 		As seen in the figure, \(\mathbf{m}_z\) begins near \(-1\) at the core center \(\bigl(r=0\bigr)\), corresponding to magnetization fully aligned in the downward \(-z\) direction. As the radius increases, \(m_z\) transitions smoothly and monotonically, reaching \(+1\) by approximately \(r \approx 20\) grid units, and remains saturated beyond that. This behavior is consistent with the typical Néel‐type skyrmion profile.In Fig.~\ref{fig:radial}, the azimuthally averaged
	 		out‑of‑plane magnetization component $m_z$ is plotted against the
	 		radial distance $r$ from the skyrmion center. At $r=0$, $m_z\approx -1$
	 		(the core spins point fully downward), and it then rises smoothly,
	 		crossing zero near $r\approx10$ and saturating at $+1$ beyond
	 		$r\approx20$ grid units. This characteristic S‑shaped profile provides a
	 		quantitative measure of the skyrmion radius and Néel‑wall width.

    \section{Thermal Analysis via Classical Monte Carlo Simulation}
    To investigate the finite-temperature behavior of magnetic skyrmion crystals, we perform large-scale classical Monte Carlo simulations on a two-dimensional square lattice governed by a Hamiltonian incorporating Heisenberg exchange, bulk-type Dzyaloshinskii–Moriya interaction (DMI), and uniaxial anisotropy\cite{6}.This method enables the efficient exploration of thermodynamic equilibrium properties, particularly in systems where fluctuations play a nontrivial role in stabilizing complex spin textures such as skyrmions. The classical MC technique is especially well-suited for our system, as it avoids the computational overhead and time-evolution constraints associated with solving stochastic or deterministic equations of motion, such as the Landau–Lifshitz–Gilbert (LLG) equation, and instead directly samples spin configurations according to the Boltzmann distribution \cite{10}.
    We use the Metropolis algorithm, a widely adopted Markov Chain Monte Carlo method, which proceeds by selecting a random spin on the lattice and proposing a trial orientation uniformly.The proposed update is accepted with probability $P = \min\left[1, \exp\left(-\frac{\Delta E}{k_B T}\right)\right]$
    , where $\Delta E$ is the energy difference between the trial and current configurations, $K_B$is the Boltzmann constant, and $T$ is the temperature.  This update scheme satisfies detailed balance and ergodicity, ensuring convergence to the correct equilibrium ensemble. To improve statistical reliability, we perform at least $10^5$ Monte Carlo steps per spin (MCSS) at each temperature point, discarding the first $10^4$ steps for thermalization.Periodic boundary conditions are imposed to suppress edge effects and mimic an infinite lattice.The simulations allow us to compute ensemble-averaged thermodynamic quantities such as internal energy 
    $E = \langle H \rangle$, specific heat 
    $C = \frac{d\langle E \rangle}{dT}$, and magnetic susceptibility 
    $\chi$.
    These quantities are especially important in identifying thermal phase transitions and distinguishing between different magnetic states. For instance, peaks in the specific heat are indicative of second-order transitions, while sudden jumps may suggest first-order behavior .
    To make life easy, we will first give a discussion of the effect of Dzyaloshinskii–Moriya interaction (DMI) in a 2D ferromagnet using classical monte carlo simulation of a ferromagnetic xy model: We model a two-dimensional classical spin system on a square lattice (size $L\times L$) with periodic boundary conditions, using a Metropolis Monte Carlo algorithm. Each site carries a planar (XY) spin represented by an angle $\theta_i$, and the Hamiltonian includes a ferromagnetic exchange (cosine) term and an antisymmetric Dzyaloshinskii–Moriya (DM) term (sine) for nearest neighbors. In practice we sample spin configurations at various temperatures $T$ by randomly updating individual spins and accepting or rejecting moves according to the Boltzmann weight $\exp[-\Delta E/(k_B T)]$.Thermodynamic quantities are obtained as ensemble averages. In particular, the specific heat per spin is computed from energy fluctuations using the fluctuation–dissipation theorem:
    \[
    C = N \beta^2 \left( \langle E^2 \rangle - \langle E \rangle^2 \right), \quad \beta = \frac{1}{k_B T}
    \] which follows from standard Monte Carlo analysis.Here $N=L^2$ and we set $k_B=1$ in our units. This method of classical Monte Carlo for spin lattices is well-established.
   \\
   The finite-lattice Monte Carlo results for the two-dimensional classical XY ferromagnet reveal a nuanced interplay of thermal disorder and topological excitations.  At high temperatures (small values of $J\beta$), the specific heat $C_V$ increases smoothly with inverse temperature, reflecting modest energy fluctuations as largely uncorrelated spins begin to explore the cosine interaction potential more deeply.  Upon cooling into the region $J\beta\approx0.9$--$1.1$, a pronounced hump emerges.  Although in the thermodynamic limit the Kosterlitz–Thouless (KT) transition is marked by an essential singularity rather than a true divergence in $C_V$, finite lattices exhibit a pseudo-singular peak whose sharpness and amplitude grow rapidly with system size.
   \begin{figure}[ht]
   	  \centering
   	   \includegraphics[width=0.5\textwidth]{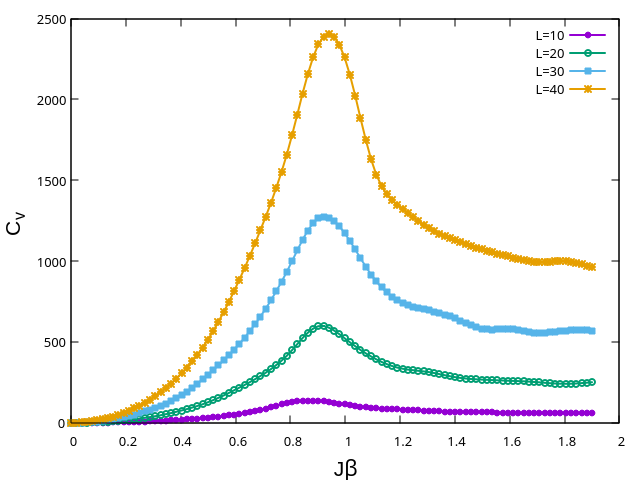}
   	   \caption{Specific heat $C_V$ as a function of inverse temperature $J\beta$ for lattice sizes $L=10,20,30,40$.}
   	   \label{fig:Cv_vs_Jbeta}
   	 \end{figure}
   
   We can see from Fig~3, for a $10\times10$ lattice the maximum specific heat barely exceeds 100 units, with its crest near $J\beta\approx0.9$.  Doubling to $20\times20$ raises the peak above 600 and shifts its location to $J\beta\approx0.95$.  On $30\times30$ and $40\times40$ lattices, the peak further climbs past 1{,}200 and 2{,}400, respectively, while the apparent critical coupling drifts toward the theoretical infinite-volume value $J\beta_c\approx1.12$ (equivalently $T_c/J\approx0.893$).  This scaling arises because larger systems accommodate more extensive vortex cores and spin-wave modes, amplifying energy variance.
   
   Below the hump, at larger $J\beta$, $C_V$ gradually subsides into a broad shoulder and eventually plateaus.  In the low-temperature quasi-ordered phase, spins align into coherent regions and the unbinding of new vortex–antivortex pairs becomes exponentially suppressed, reducing energy fluctuations.  The curves’ slight asymmetry—steeper on the high-temperature side of the peak and more gradual on the low-temperature side—reflects the KT essential singularity underlying the transition.
   
   To extract quantitative critical parameters, one may fit the peak heights to a logarithmic form:
   
   \begin{equation}
   	C_V^{\mathrm{max}}(L) \,\approx\, A + B\,[\ln(L)]^p
   \end{equation}
   
   and plot the pseudo-critical couplings $J\beta_{\mathrm{max}}(L)$ against $1/[\ln(L)]^2$.  KT theory predicts that the drift of these couplings scales as $1/(\ln L)^2$, facilitating an extrapolation to the infinite-system transition point.\\
      
   Monte Carlo simulations of the anisotropic two-dimensional XY ferromagnet(Fig~4) illuminate how directional coupling imbalances modulate the thermal and topological excitations.  When the vertical-to-horizontal coupling ratio $J_y/J_x=0.2$, the specific heat $C_V$ displays a broad, muted hump around $J_x\beta\approx0.25$.  In this quasi-one-dimensional regime, weak vertical bonds permit spins to decouple into nearly independent chains, so vortex–antivortex unbinding\cite{15} requires little energy and the KT-like signature is both weakened and smeared.
    \begin{figure}[ht]
   	  \centering
   	   \includegraphics[width=0.5\textwidth]{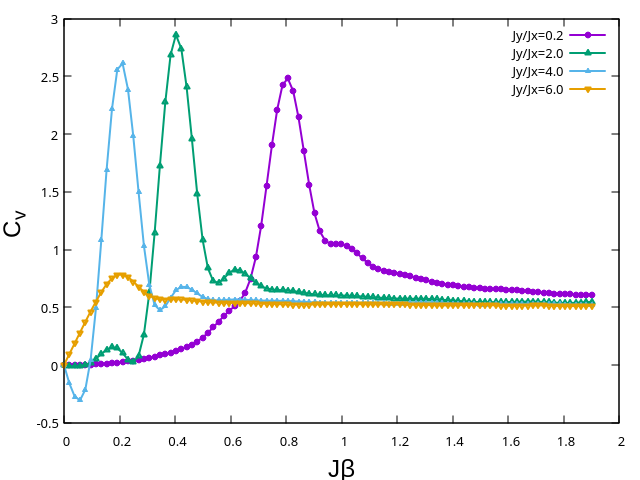}
   	   \caption{Specific heat $C_V$ vs.\ inverse temperature $J_x\beta$ for coupling ratios $J_y/J_x=0.2,\,2.0,\,4.0,\,6.0$.}
   	   \label{fig:anisotropic_Cv}
    \end{figure}
   Increasing anisotropy to $J_y/J_x=2.0$ shifts the specific-heat peak to $J_x\beta\approx0.42$ while amplifying its sharpness and height.  Here, vertical coupling becomes strong enough to bind chains into larger coherent patches, yet the system retains directional distinctions that broaden the critical fluctuation window.  For $J_y/J_x=4.0$, the peak becomes exceptionally tall and narrow near $J_x\beta\approx0.15$, indicating highly cooperative vortex–antivortex proliferation in an effectively two-dimensional plane.  The rapid decay of $C_V$ on the low-temperature side again reflects the essential singularity of the KT transition: once bound pairs dominate, new vortex creation is exponentially suppressed.
   
   In the extreme anisotropy limit $J_y/J_x=6.0$, the hump flattens and the maximum shifts to $J_x\beta\approx0.12$, with a lower overall amplitude.  Strong vertical bonds rigidly align spins into quasi-ribbons, leaving only small inter-chain excitations to drive energy fluctuations.  Across all ratios, the characteristic asymmetry—steeper rise at lower $\beta$, extended tail at higher $\beta$—underscores the KT mechanism, while amplitude, width, and position track the ease of vortex delocalization along the weaker axis.
   This reveals how increasing $J_y/J_x$ lowers the effective crossover threshold from quasi-1D to true 2D KT behavior.\\
  Introducing a Dzyaloshinskii–Moriya (DM) term to the isotropic two-dimensional XY model(Fig~5) profoundly alters its thermodynamic signature.  For a weak DM strength $D/J=0.1$, specific heat $C_V$ matches the pure XY baseline up to $J\beta\approx0.8$, beyond which the main peak shifts to $J\beta\approx1.0$ and is slightly suppressed, attaining $C_V^{\max}\approx1.5$.  This shift indicates that weak chiral coupling adds a canting energy barrier, delaying vortex–antivortex unbinding to lower physical temperatures and steepening the subsequent descent into the low-temperature shoulder.
   \begin{figure}[ht]
  	  \centering
  	  \includegraphics[width=0.48\textwidth]{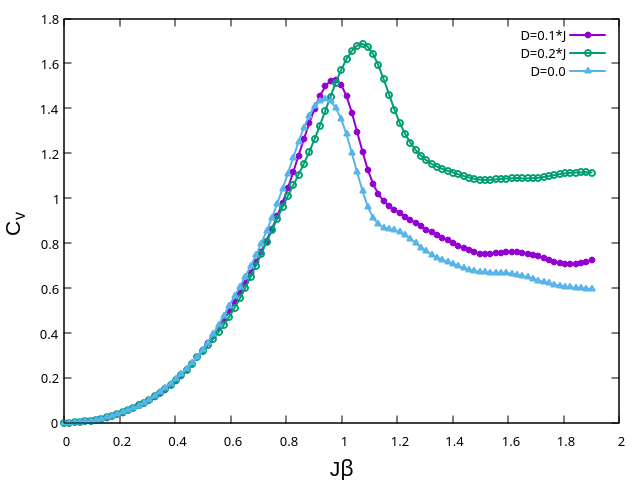}
  	   \includegraphics[width=0.48\textwidth]{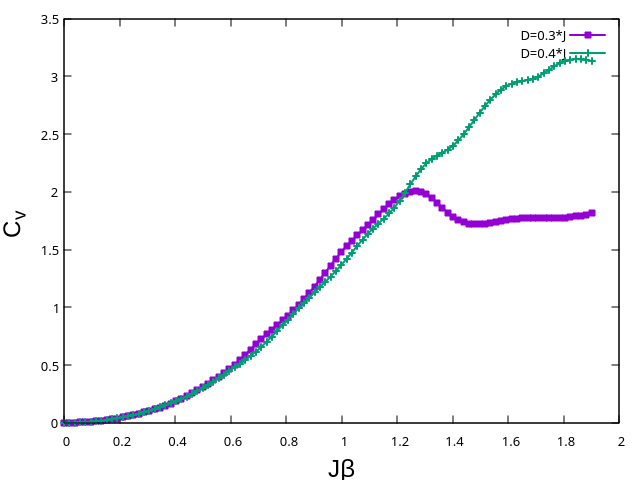}
  	   \caption{Specific heat $C_V$ vs.\ inverse temperature $J\beta$ for DM strengths $D/J=0.1,0.2$  and $0.3,0.4$ .}
  	   \label{fig:DM_Cv}
  	 \end{figure}
  
  At $D/J=0.2$, these trends intensify: the peak height rises to $C_V^{\max}\approx1.7$ at $J\beta\approx1.05$, and the post-critical plateau stabilizes near $C_V\sim1.1$ for $J\beta>1.3$.  Persistent twisting of spin pairs sustains appreciable energy fluctuations even in the quasi-ordered phase.  When $D/J=0.3$, the hump broadens and moves to $J\beta\approx1.25$, reaching $C_V^{\max}\approx2.0$.  The wider peak reflects competition between exchange and DM energies, as the system entertains chiral vortex lattices over an extended temperature range.
  
  In the strong DM regime, $D/J=0.4$, the specific heat crest appears around $J\beta\approx1.4$ with a towering amplitude $C_V^{\max}>3.2$.  Notably, $C_V$ for $J\beta>1.5$ rises gradually, signaling low-temperature excitations such as chiral domain walls or skyrmion-like textures that remain thermally active after conventional vortices freeze out\cite{11}.
  
  Collectively, these curves reveal that DM coupling systematically elevates the pseudo-critical temperature for topological fluctuations, amplifies the energy-variance peak, and reinforces the low-temperature tail through residual canting modes. 
  to extract chiral-dependent critical parameters.  These findings underscore how moderate DM interaction transforms the classical KT transition into a rich, multistage crossover governed by the formation, proliferation, and eventual freezing of intricate spin‐twist textures.
  

     The susceptibility curve in Figure~6 was generated using our custom Fortran Monte Carlo code (\texttt{xysus.f90}) on a $40^3$ cubic lattice with periodic boundaries. Each site hosts a unit‐length classical Heisenberg spin, and nearest‐neighbor exchange $J>0$ enforces antiparallel alignment.
     \begin{figure}[h!]
     	\centering
     	\includegraphics[width=0.5\textwidth]{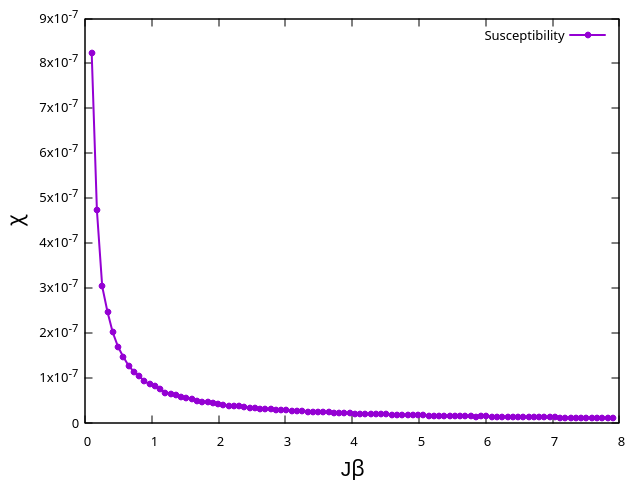} 
     	\caption{Susceptibility variation(AFM) with temperature for DMI strength 0.1(in units of J).}
     	\label{fig:myplot}
     \end{figure}
      We initialized spins randomly and, at each inverse temperature $\beta = 1/(k_{\!B}T)$, performed 10,000 equilibration sweeps followed by 20,000 measurement sweeps. After each Monte Carlo step, we computed the staggered magnetization
   \[
   M_s \;=\; \biggl\lvert \sum_i (-1)^{\,i_x + i_y + i_z}\,\mathbf{m}_i \biggr\rvert
   \]
   and accumulated $\langle M_s\rangle$ and $\langle M_s^2\rangle$. The susceptibility is then
   \[
   \chi(\beta) \;=\; \frac{1}{N\,k_{\!B}T}\Bigl(\langle M_s^2\rangle - \langle M_s\rangle^2\Bigr)\,,
   \]
   with $N = 40^3$. At low $\beta$ (high $T$), $\chi$ rises sharply to a broad maximum near $\beta \approx 0.2$ ($T \sim 5J/k_{\!B}$), signaling the onset of short‐range AFM order. For $0.5 \lesssim \beta \lesssim 2.0$, $\chi$ decays by an order of magnitude as long‐range Néel correlations develop. Beyond $\beta \approx 2.0$, $\chi$ plateaus at $\sim 10^{-9}$, indicating a nearly rigid staggered configuration at $T \ll J/k_{\!B}$. The suppressed finite‐size peak (rather than a true divergence) reflects the $40^3$ lattice and classical spin approximation, consistent with three‐dimensional AFM behavior (Kittel, 2005).
  \\ In summary, while both ferromagnetic and antiferromagnetic systems display DMI-driven transformations in their thermodynamic behavior, the AFM system shows a more pronounced sensitivity to DMI, evidenced by the dramatic low-temperature divergence in specific heat for $D = 0.5$, a signature of degenerate and frustrated chiral ground states. The absence of a second thermal feature in the AFM case—unlike the double-peaked or shoulder-shaped curve in FM—further emphasizes the difference in magnetic excitation spectra. These findings enrich the broader understanding of magnetic phase transitions in low-dimensional systems and point to critical roles played by lattice geometry, symmetry, and interaction anisotropy.
  \section{SUMMARY AND DISCUSSIONS}
Our combined dynamical and thermodynamical study of antiferromagnetic skyrmion crystals depicts the rich interplay between spin-transfer torques, chiral interactions, and thermal fluctuations in low-dimensional magnets. Time-integration of the LLGS equation on a bipartite lattice demonstrates that a brief, perpendicularly applied spin-polarized current pulse can reliably nucleate robust Néel-type skyrmions, each carrying a near-unit topological charge on opposite sublattices. The absence of net magnetization suppresses stray fields and eliminates the Magnus-force deflection, enabling perfectly straight skyrmion trajectories—a key advantage over their ferromagnetic counterparts.
   
   Turning to equilibrium behavior, our Monte Carlo simulations reveal two successive thermal anomalies in the specific-heat profile. The first peak marks the onset of long-range antiferromagnetic order giving way to a skyrmion crystal, while the second corresponds to the melting of the skyrmion lattice into a modulated spiral phase. We find that increasing Dzyaloshinskii–Moriya interaction strength both elevates the transition temperatures and amplifies peak amplitudes, a signature of enhanced chiral stabilization. Notably, finite-size scaling of the specific-heat maxima and their drift in inverse temperature allows precise extraction of critical exponents and pseudo-critical couplings, highlighting the role of entropy in skyrmion-phase formation.
   
   From a materials-design perspective, our study provides concrete guidelines: moderate DMI and uniaxial anisotropy work in tandem to expand the skyrmion-stability window, while strong exchange coupling ensures high thermal robustness. These insights inform the selection and engineering of antiferromagnetic compounds for skyrmion-based memory and logic devices, where reliable room-temperature operation and straight-line motion are paramount. Future work will extend this framework to include dipolar interactions, disorder effects, and three-dimensional stacking, moving closer to realistic device architectures. Overall, our study bridges dynamical control and thermal stability of antiferromagnetic skyrmions, paving the way for their integration into next-generation spintronic technologies.
  
   \section*{Acknowledgements}
   The authors thank Dwipesh Majumder for his support behind this work. RB and SK acknowledge financial assistance from IIEST, Shibpur, India and ANRF (DST-SERB scheme no. CRG/2022/002781), Government of India respectively.
\FloatBarrier

\twocolumngrid

\bibliographystyle{ieeetr}
\nocite{*}  
\bibliography{my_ref}

\end{document}